\documentclass[pra,aps,reprint,citeautoscript,amsmath,amssymb,superscriptaddress]{revtex4-1}
\usepackage{graphicx}
\usepackage{dcolumn}
\usepackage{bm}
\usepackage{amsmath}
\usepackage{hyperref} 
\usepackage{soul} 
\DeclareGraphicsExtensions{.pdf,.eps,.png,.jpg,.mps}
\usepackage{mathrsfs}
\usepackage{soul} 
\usepackage[utf8]{inputenc} 
\usepackage{hyperref} 
\usepackage{xcolor} 
\usepackage{amsmath}
\usepackage{amsthm}
\usepackage{eucal}
\usepackage{amssymb}

\usepackage{booktabs}

\begin{document}
\title{Gate-tunable direct and inverse electrocaloric effect in trilayer graphene}

\author{Natalia Cortés} 
\email{natalia.cortesm@usm.cl}
\affiliation{Departamento de Física, Universidad Técnica Federico Santa María, Casilla 110V, Valparaíso, Chile}
\author{Oscar Negrete}
\affiliation{Departamento de Física, Universidad Técnica Federico Santa María, Casilla 110V, Valparaíso, Chile}
\affiliation{Centro para el Desarrollo de la Nanociencia y la Nanotecnolog\'ia, 8320000 Santiago, Chile}
\author{Francisco J. Peña}
\affiliation{Departamento de Física, Universidad Técnica Federico Santa María, Casilla 110V, Valparaíso, Chile}
\author{Patricio Vargas}
\affiliation{Departamento de Física, Universidad Técnica Federico Santa María, Casilla 110V, Valparaíso, Chile}
\affiliation{Centro para el Desarrollo de la Nanociencia y la Nanotecnolog\'ia, 8320000 Santiago, Chile}

\date{\today}

\begin{abstract}
The electrocaloric (EC) effect is the reversible change in temperature and/or entropy of a material when it is subjected to an adiabatic electric field change. Our tight-binding calculations linked to Fermi statistics, show that the EC effect is sensitive to the stacking arrangement in trilayer graphene (TLG) structures connected to a heat source, and is produced by changes of the electronic density of states (DOS) near the Fermi level when external gate fields are applied on the outer graphene layers. We demonstrate the AAA-stacked TLG presents an inverse EC response (cooling), whereas the EC effect in ABC-stacked TLG remains direct (heating) regardless of the applied gate field potential strength. We reveal otherwise the TLG with Bernal-ABA stacking geometry generates both the inverse and direct EC response in the same sample, associated with a gate-dependent electronic entropy transition at finite temperature. By varying the chemical potential to different Fermi levels, we find maxima and minima of the DOS are located near the extremes of the electronic entropy, which are correlated with sign changes in the differential entropy per particle, giving a particular experimentally measurable electronic entropy spectrum for each TLG geometry. The EC effect in quantum two-dimensional layered systems may bring a wide variety of prototype van der Waals materials that could be used as versatile platforms to controlling the temperature in nanoscale electronic devices required in modern portable on-chip technologies. 
\end{abstract} 


\maketitle
\date{Today} 
\section{Introduction}
The rapid increase in the need for efficient, environmentally friendly and low-cost materials with the capability of cooling, has opened interesting possibilities to explore material properties beyond the traditional vapor-compression method applied in household and industrial refrigeration \cite{scott2011electrocaloric,shi2019electrocaloric}. As a clean and efficient alternative to tuning the systems temperature, can be the implementation of caloric effects on solid-state materials, driven by applied pressure, electrical, magnetic and/or mechanical fields, conducting to the barocaloric, electrocaloric (EC), magnetocaloric and elastocaloric effect, respectively \cite{moya2014caloric,shi2019electrocaloric,li2019colossal}. Thermal response on solid systems is produced by either of those external field acting on and/or removed from the sample under adiabatic conditions \cite{valant2012electrocaloric}, leading to large temperature changes $\Delta T$ associated with isothermal entropy changes $\Delta S_{T}$, generally near phase transitions \cite{moya2014caloric,otonivcar2020electrocalorics}.

In the last decades, the EC effect---the change in temperature and entropy as an electric field is applied---has been widely studied in three-dimensional dielectric multilayer capacitors (MLC), by considering a different number of layers with diverse thicknesses in the samples, typically of the order of micrometers \cite{kar2010direct,moya2018electrocaloric}, and finding thickness-dependent thermal responses \cite{moya2018electrocaloric}. In single-crystal MLC, the temperature change reaches $\sim0.9$ K near ferroelectric phase transitions when moderate electric fields are applied \cite{moya2013giant}. In high-quality samples of MLC, the temperature change increases up to 5.5 K \cite{nair2019large,otonivcar2020electrocalorics}. 
As films of MLC scales down to hundreds of nanometers, a giant EC response of $\sim12$ K is manifested close phase transitions \cite{mischenko2006giant}. Theory predicts larger cooling power in thin-film MLC of ceramics and polymers by varying the number of layers of the systems \cite{kar2009predicted}. The feasible scalability of MLC systems to reduced dimensionality may allow expanding the EC effect to novel materials for electrical refrigeration at the atomistic level \cite{lisenkov2009intrinsic,ponomareva2012bridging}. 

Nanoscale systems including van der Waals multilayers and heterostructures, comprise a wide variety of quantum solids \cite{novoselov2016} that could be used as prototype miniaturized materials to get caloric effects. A giant EC response of $\sim23$ K has experimentally been achieved in a ferroelectric heterobilayer because of interface-induced interactions  \cite{shirsath2019interface}. Theoretical studies on graphene nanoribbons under an electric and magnetic field show controllable entropy changes magnitude and entropy sign changes due to the applied magnetic field \cite{reis2013electrocaloric}. First-principle and thermodynamic calculations in a two-dimensional (2D) monolayer of MoTe$_2$, have reported a temperature change of $10-15$ K near the structural phase transition that occurs when the monolayer is subject to electrostatic gating \cite{rehn2018refrigeration}. 

We explore here the thermal response of atomically thin layered nanostructures, focusing on gated trilayer graphene (TLG). Both experimental and theoretical studies have revealed the electronic band structure and density of states (DOS) of TLG samples strongly depend on the stacking pattern they possess, showing significant changes under applied electrical potentials \cite{aoki2007dependence,avetisyan2009electric,koshino2009gate,avetisyan2010stacking,craciun2009trilayer,lui2011observation,yankowitz2013local,zou2013transport,wang2016first}. This makes TLG materials suitable to inspect the EC effect and novel thermodynamic quantities such as the differential entropy per particle (DEP) \cite{kuntsevich2015strongly,tsaran2017entropy,grassano2018detection,shubnyi2018entropy,galperin2018entropy,sukhenko2020differential}. Through Bloch electrons in parametrized tight-binding models linked to Fermi statistics, we show how the thermal response varies within each TLG stacking as a result of the quantum-thermodynamic processes involved at low-energies. As each system is gated, we find entropy changes for AAA-stacked TLG linearly increase with temperature, showing an inverse EC effect (cooling) up to room temperature. In contrast, the entropy changes for ABC TLG non-linearly increase as temperature increases, giving a direct EC effect (heating). Remarkably, the ABA-TLG arrangement presents a dual EC response as cools and heats within the same sample, displaying gate-dependent maxima for entropy changes. We also analyze the number of electrons thermally excited of ABA-TLG at low and high temperatures, unveiling its direct connection to the gate-tunable electronic entropy and dual EC effect.   

\section{Quantum-Thermodynamic Model}
\begin{figure*}[ht!]
\centering
\includegraphics[width=\linewidth]{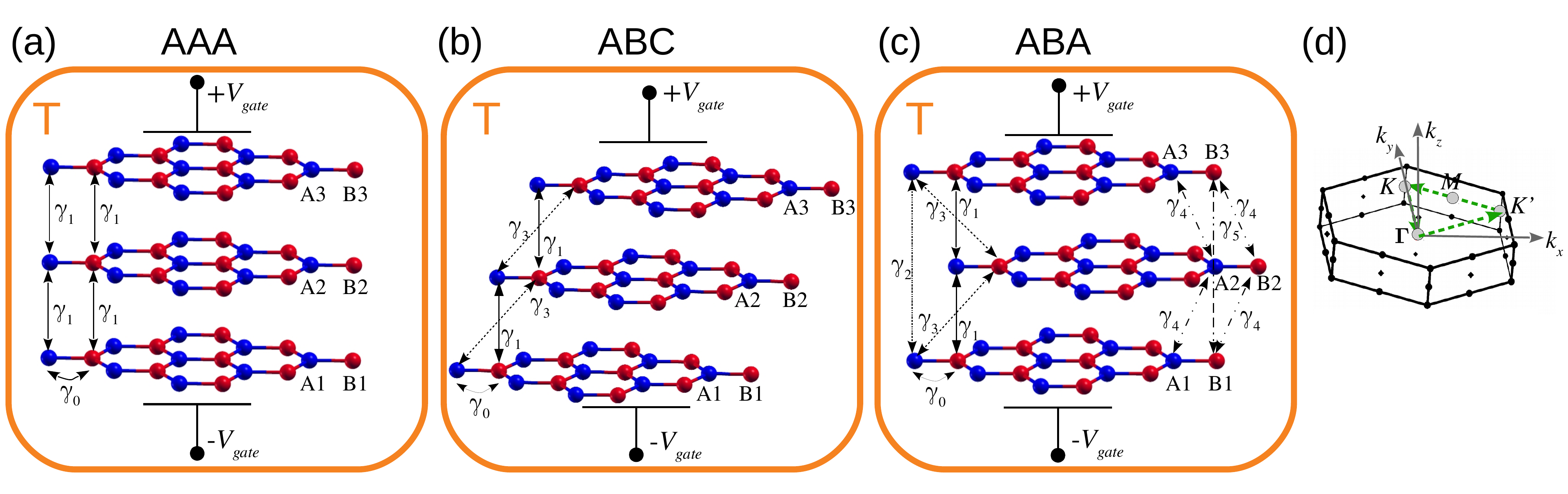}
\caption{Schematic representation of trilayers graphene with diverse stacking patterns, hexagonal AAA (a), rhombohedral ABC (b) and Bernal ABA (c). Blue (red) sphere show $A$ ($B$) carbon sublattice connected through intralayer $\gamma_0$ and interlayer hoppings $\gamma_1$, $\gamma_2$, $\gamma_3$, $\gamma_4$ and $\gamma_5$ as indicated; $\pm V_{gate}$ is the gate potential applied to the outer graphene layers on each system, the orange square enclosing each structure represent the thermal source at temperature $T$. (d) Shows the Brillouin zone in momentum space for all trilayers graphene, where green lines highlight the triangular $\boldsymbol{k}$-path with corners located at high symmetry points $\Gamma=(0,0)$, $K=\frac{2\pi}{a}\big(0,\frac{2}{3}\big)$ and $K'=\frac{2\pi}{a}\big(\frac{\sqrt3}{3},\frac{1}{3}\big)$, whit $a=\sqrt{3}a_{\text{C-C}}=2.46$ \AA\ the 2D graphene lattice constant.}\label{fig1}
\end{figure*}

We simulate TLG structures with simple hexagonal AAA, Bernal ABA, and rhombohedral ABC stacking order by using a $\pi$-orbital tight-binding (TB) model, that captures the main low-energy symmetry properties around the Fermi level for each system \cite{partoens2006graphene,partoens2007normal,koshino2009trigonal,avetisyan2009electric,koshino2009gate,avetisyan2010stacking,yuan2010electronic,mohammadi2014effects}. To construct the TB Hamiltonians, we consider the unit cells of the TLG with three pairs of atoms \{$A1$-$B1$\}, \{$A2$-$B2$\} and \{$A3$-$B3$\}, which are located respectively in the bottom, central and top graphene layers, as shown in Fig.\ \ref{fig1}. We use different parametrization (hoppings) depending on the stacking type of the TLG. For each system we have first taken a minimal set of parameters from fitting photoemission spectroscopy spectra to TB models \cite{bao2017stacking}. These parameters include a strong intralayer hopping $\gamma_0$ between $Ai$ and $Bi$ nearest neighbors within each graphene layer ($i=1,2,3$ is the layer index), and one interlayer hopping connecting direct vertical nearest-layer carbon sites with strong coupling $\gamma_1$, as used for AAA-stacked TLG in Fig.\ \ref{fig1}(a). For ABC and ABA-stacked TLG, this parametrization also incorporates a weaker nearest-layer coupling $\gamma_3$, Fig.\ \ref{fig1}(b) and  \ref{fig1}(c) respectively. In particular, for ABA-TLG, we also have taken a full set of parameters adopting the Slonczewski-Weiss-McClure parametrization \cite{dresselhaus2002intercalation}, which contains one extra nearest-interlayer hopping $\gamma_4$ as well as two next-nearest interlayer hoppings $\gamma_2$ and $\gamma_5$. An asymmetric onsite-potential energy difference with magnitude $2V_{g}$, induced by a bias voltage between the external graphene layers, can act as an external gate potential on each carbon atom of top $+V_{g}$ and bottom $-V_{g}$ graphene layers of the TLG structures, similar to bilayer graphene under an external voltage \cite{hao2010thermopower}. This asymmetric potential is a crucial quantity to tune the electronic DOS and to produce a caloric response near the Fermi level of TLG systems.

The Hamiltonians for the TLG samples $\mathcal{\hat{H}}^{\alpha}(\pmb{k})=\hat{H}^{\alpha}_{0}(\pmb{k})+\hat{V}$ ($\alpha=\mathrm{AAA,ABA,ABC}$) include a diagonal matrix $\hat{V}$ of the gate potential proportional to $V_g$. The matrices of hopping $\hat{H}^{\alpha}_{0}(\pmb{k})$ can be obtained with the matrix elements $H_{jj'}^{\alpha}(\pmb{k})=\sum_{\pmb{R}} e^{i\pmb{k}\cdot \pmb{R}} E_{jj'}^{\alpha}(\pmb{R})$, where $E_{jj'}^{\alpha}(\pmb{R})=\left\langle\phi_{j}(\pmb{r})\right|\hat{H}_{0}^{\alpha}\left|\phi_{j'}(\pmb{r-R})\right\rangle$ is the hopping integral between the atomic orbitals $\left|\phi_{j}\right\rangle$ at $\textbf{0}$ and $\left|\phi_{j'}\right\rangle$ at lattice vector $\pmb{R}$ with $j(j')=Ai,Bi$. $\pmb{R}_1=a_{\text{C-C}}(1,0)$, $\pmb{R}_{2(3)}=a_{\text{C-C}}\big(-1/2,+(-)\sqrt{3}/2 \big)$ are in-plane nearest-neighbor vectors with $a_{\text{C-C}}=1.42$ \AA\ the carbon-carbon distance within a graphene layer, and $\pmb{k}=(k_x, k_y, k_z$) is the momentum. The TLG Hamiltonians are represented in the basis with components $\{\psi_{A1},\psi_{B1},\psi_{A2},\psi_{B2},\psi_{A3},\psi_{B3}\}$.

The density of states $D(E,V_g)$ is a fundamental gate-tunable electronic quantity, which allows us to calculate the electronic thermodynamics properties of the TLG systems. The DOS depends of both, the electronic level with energy $E$ and the gate potential $V_g$ applied on the outer carbon atoms of each TLG. As the TLG structures can be considered as quasi 2D systems, we numerically calculate the DOS using a 2D Brillouin zone (BZ) in reciprocal $\pmb{k}$-space (i. e., $k_z=0$). We use a fine mesh of about ten million of $\pmb{k}$ points in the area enclosed by the green triangle of Fig.\ \ref{fig1}(d), and for every $\pmb{k}$-state we evaluate the energy levels coming from each band of $\mathcal{\hat{H}}^{\alpha}(\pmb{k})$. By calculating the DOS, we can obtain the number of electrons $N$ in each unit cell of the trilayers
\begin{equation}\label{numelectrons}
N(V_g,T,\mu)=\int_{E_l}^{E_h}D(E,V_g)n_\text{F}(E,T,\mu)dE,
\end{equation}
where $E_{l(h)}$ is the lowest (highest) electronic energy eigenvalue of the considered Hamiltonian, $n{_\text{F}}(E,T,\mu)=1/[e^{\beta(E-\mu)}+1]$ is the Fermi-Dirac function distribution with $\beta=1/k_{\text{B}}T$, $k_{\text{B}}$ is the Boltzmann constant, $\mu$ is the chemical potential and $T$ temperature. At $T=0$ K, the electronic levels are occupied up to the Fermi energy $E_{\text{F}}$, and $n{_\text{F}}$ it converts in the Heaviside function, so that we obtain $N=\int_{E_l}^{E_{\text{F}}}D(E,V_g)dE=6$ for the unit cell of each TLG. For finite temperatures $T>0$, we can calculate the chemical potential $\mu(V_g,T)$ by inversion of Eq.\ \ref{numelectrons}. 

The total entropy of the system $S=S_{\text{latt}}(T)+S_{e}(V_g,T)$ includes two terms, the entropy of the lattice $S_{\text{latt}}(T)$, giving account of the phonon contribution, in which we assume it is only dependent on temperature and not on $V_{g}$ \cite{kutnjak1999electrocaloric}. In our approximation we neglect $S_{\text{latt}}(T)$ and use $S \sim S_{e}(V_g,T)$, where the electronic entropy 
\begin{equation}\label{entropy}
S_{e}(V_g,T)=-k_{\text{B}} \int_{E_l}^{E_h}D(E,V_g)\mathcal{F}(n_\text{F})dE,
\end{equation}
depends on the gate potential and temperature, and is calculated in the triangular area of the BZ of Fig.\ \ref{fig1}(d), where the number of electrons is fixed to $N=6$. In Eq.\ \ref{entropy},
\begin{equation}\label{funcionnumero}
    \mathcal{F}(n_\text{F})=n_\text{F}\ln n_\text{F}+(1-n_\text{F})\ln(1-n_\text{F}), 
\end{equation}
its approximated by a Lorentzian-like function $L(E,T,\mu)=C/[e^{(|E-\mu|/2k_{\text{B}}T)^{3/2}}+1]$. By considering $C=1.4$, low and high $T$ values, we obtain excellent agreement between Eq.\ \ref{funcionnumero} and $L(E,T,\mu)$ with $-\mathcal{F}(n_\text{F})\approx L(E,T,\mu)$, so that Eq.\ \ref{entropy} transforms as
\begin{equation}\label{aproxentropy}
S_{e}(V_g,T)\simeq k_{\text{B}} \int_{E_l}^{E_h}D(E,V_g)L(E,T,\mu)dE.
\end{equation} 
The main contribution of $L(E,T,\mu)$ to the electronic entropy is given by their temperature-dependent width, which increases as temperature increases. Within the range of temperature we work, $k_\text{B}T \ll \gamma_1$, the chemical potential remains constant at $\mu=0$ eV in order to fulfill $N=6$ for the unit cell of each
TLG. 

The EC effect is in general quantified by either the temperature change that occurs when the electric field changes adiabatically, or the entropy change induced by isothermal application or removal of the electric field. Typically there are two ways to measure the caloric effects in solid materials: (\textit{i}) through the variation of temperature in an adiabatic thermodynamic path $\Delta T$ (direct measurement), and (\textit{ii}) by means of the entropy change in an isothermal path $\Delta S_{T}$ (indirect measurement), where the subindex $T$ indicates constant temperature. It is important to note that is experimentally challenging to obtain $\Delta T$ as compared to $\Delta S_{T}$, as direct measurements generally require precision calorimetry \cite{scott2011electrocaloric}.

The EC effect in our TLG structures is then obtained through the indirect way, by means of electronic entropy changes calculations as a function of $T$ through Eq.\ \ref{aproxentropy} at $\mu=E_{\text{F}}=0$,
\begin{equation}
\label{deltaS}
-\Delta S_{e,T}(T)= S_{e}(V_{g}=0,T)-S_{e}(V_{g},T).
\end{equation}
In case to obtain $-\Delta S_{e,T}>0$, we are in presence of the direct EC effect, that is the TLG system is capable to heat as $V_g$ increases. In the opposite case when $-\Delta S_{e,T}<0$, the system present an inverse EC effect and so the sample cools down. 

\subsection{Differential electronic entropy per particle}\label{sec:DEP}
To get essential insights into the microscopic origin of the EC effect in TLG structures, we have calculated a fundamental thermodynamic quantity, the differential electronic entropy per particle $s$ (DEP). The DEP is an experimentally measurable quantity in gated 2D electron systems, allowing superior sensitivity in comparison to a.c. calorimetry \cite{kuntsevich2015strongly}, and can be directly linked to the thermoelectric power or alternatively to the Seebeck coefficient \cite{goupil2011thermodynamics}. The DEP, $s=\partial S_{e}/\partial N$ is found from the Maxwell relation as \cite{kuntsevich2015strongly,tsaran2017entropy,grassano2018detection,shubnyi2018entropy,galperin2018entropy,sukhenko2020differential}
\begin{equation}\label{DEP}
s(\mu,V_g,T)=\left(\frac{\partial S_{e}}{\partial N}\right)_{T}=-\left(\frac{\partial \mu}{\partial T} \right)_{N},
\end{equation} 
and its related to the DOS through $N$ in Eq.\ \ref{numelectrons} by means $s(\mu,V_g,T)=(\partial N/ \partial T)_{\mu}(\partial N/ \partial \mu)^{-1}_{T}$.

To interpret our results, we make a useful transformation of the form $s(\mu,V_g,T)=(\partial S_e/\partial N)_{T}=(\partial S_e/\partial \mu)_T(\partial \mu/\partial N)_T$. The function that predominantly governs the DEP corresponds to the entropy's derivative with respect to $\mu$. Consequently, we expect that at maxima and minima of the electronic entropy as a function of $\mu$, the DEP vanishes, giving account of the thermodynamic fingerprints for each TLG arrangement. 

As the number of particles $N$ can vary in a doped/gated system, we obtain a caloric effect directly connected to the DEP, which could be used as a feasible procedure to determine the entropy changes given by fluctuating charge carriers, see Appendix for more details. 
\section{The electrocaloric response in trilayers graphene}
\subsection{AAA-stacked TLG}
AAA-stacked TLG has been recently exfoliated \cite{bao2017stacking}, preserves metallic character in the presence of external electric fields \cite{wu2011field} as in the monolayer graphene case, and possesses mirror reflection symmetry with respect to the central layer. In this structure the \{$Ai$,$Bi$\} sublattices match to the nearest-neighbor layer carbon sites \{$Ai+1$,$Bi+1$\} with vertical hopping $\gamma_1$. The whole system effectively can be seen as a direct superposition of three graphene monolayers as shown in Fig.\ \ref{fig1}(a). Because of the crystal symmetry, the AAA-TLG Hamiltonian
\begin{equation}\label{HamiltonianAAA}
\mathcal{\hat{H}}^{\mathrm{AAA}}(\pmb{k})= \begin{pmatrix} 
-V_g & \gamma_0 f_{0} & \gamma{_1} f_{1} & 0 & 0 & 0 \\
\gamma_0 f_{0}^{*} & -V_g & 0  & \gamma{_1} f_{1} & 0 & 0 \\
\gamma{_1} f_{1}^{*} & 0 & 0 & \gamma_0 f_{0} & \gamma{_1} f_{1} & 0 \\
0 & \gamma{_1} f_{1}^{*} & \gamma_0 f_{0}^{*} & 0 & 0 & \gamma{_1} f_{1} \\
0 & 0 & \gamma{_1} f_{1}^{*} & 0 & V_g & \gamma_0 f_{0}\\
0 & 0 & 0 & \gamma{_1} f_{1}^{*} & \gamma_0 f_{0}^{*} & V_g
\end{pmatrix},
\end{equation}
allows analytical eigenvalues connecting monolayer-graphene modes $\varepsilon_{1,2}$ and AA-bilayer graphenelike modes $\varepsilon_{3,4,5,6}$ 
\begin{subequations}\label{eigenvaluesAAA}
\begin{align}
\varepsilon_{1,2}&=\pm \gamma_0\left|f_0\right|\text{,} \\
\varepsilon_{3,4}&=\pm \gamma_0\left|f_0\right|-\sqrt{V_g{^2}+2\gamma_{1}^{2}\left|f_1\right|^2}\text{,} \\
\varepsilon_{5,6}&=\pm \gamma_0\left|f_0\right|+\sqrt{V_g{^2}+2\gamma_{1}^{2}\left|f_1\right|^2}\text{.} 
\end{align}
\end{subequations}
$f_{0}(k_x,k_y)=e^{-ik_{x}a_{\text{C-C}}}+2\cos({\frac{\sqrt{3}}{2}k_{y}a_{\text{C-C}}})e^{ik_{x}a_{\text{C-C}}}$ and $f_{1}(k_z)=e^{i k_{z} c}$ are respectively an in-plane and out-of-plane momentum-dependent functions with $c=3.3$ \AA\ the interlayer distance \cite{aoki2007dependence}.
\begin{figure}[!h]
\centering
\includegraphics[width=\linewidth]{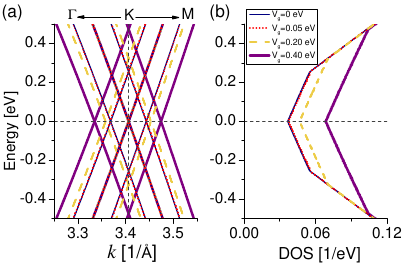}
\caption{Low-energy electronic spectra for AAA-stacked trilayer graphene. (a) Band structure in the vicinity of the $K$ point. (b) Density of states obtained for the triangular area enclosed by green lines in Fig.\ \ref{fig1}(d). Vertical dashed line in (a) highlight the $K$ point, horizontal dashed lines in (a)-(b) are fixed at $E_{\text{F}}=0$ eV. The hopping parameters are $\gamma_0=3.2$ eV and $\gamma_1=0.18$ eV \cite{bao2017stacking}.}\label{fig2}
\end{figure}
Figure \ref{fig2}(a) shows the low-energy electronic structure given by Eqs.\ \ref{eigenvaluesAAA} near the $K$ point for different values of the gate potential $0\leq V_g \sim 2\gamma_1$. Due to the structural symmetry, the Dirac points $K$,$K'$ are degenerate, where around each one of them there are three pairs of linear branches creating a diamond structure with Dirac cones throughout energy-momentum space \cite{wu2011field}. At the Fermi level $E_{\text{F}}=0$ and momentum $K$, or charge neutrality point (CNP), one of the Dirac cones remains constant regardless of the $V_g$ strength, resembling the monolayer graphene modes [Eq.\ \ref{eigenvaluesAAA}(a)]. When $V_g=0$ (thin dark blue solid lines) and because of interlayer interaction $\gamma_1$, a pair of Dirac cones  shift away from the CNP in a symmetric way by $E=\pm\sqrt{2}\gamma_{1}\simeq 0.25$ eV. As $V_g$ is turned on, a pair of cones disperse to $E=\pm\sqrt{V_g{^2}+2\gamma_{1}^{2}}$, Eq.\ \ref{eigenvaluesAAA}(b) and \ref{eigenvaluesAAA}(c). The effect of the gate potential therefore can be represented as a renormalization of the interlayer hopping energy \cite{redouani2018multibands}, preserving metallic character for the bands as in the unperturbed case ($V_g=0$). At low-energies, the DOS preserves electron-hole symmetry [$D(E,V_g)=D(-E,V_g)$] in Fig.\ \ref{fig2}(b), and shows a global minimum at the CNP, which causes the major contribution to the EC response at low and high temperatures. At the energies given by the shifted Dirac cones from the Fermi level, the DOS shows two symmetric discontinuities, not providing states to the EC effect as the $L$ function in Eq.\ \ref{aproxentropy} does not capture those states up to room temperature.  

\begin{figure}[!h]
\centering
\includegraphics[width=\linewidth]{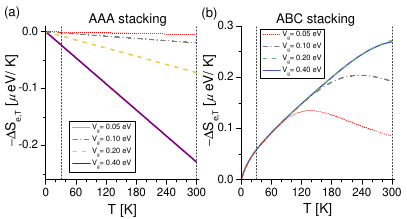}
\caption{The isothermal entropy change $-\Delta S_{e,T}$ as a function of $T$ for different gate potentials as indicated. (a) AAA-stacked and (b) ABC-stacked trilayer graphene. The chemical potential is set to $\mu=0$. Dashed vertical lines indicate $T=30$ K and $T=300$ K, where we calculate the electronic entropy and DEP in Fig.\ \ref{fig7} and Fig.\ \ref{fig8}.}\label{DeltaSAAA-ABC}
\end{figure}

\begin{figure}[!h]
\centering
\includegraphics[width=\linewidth]{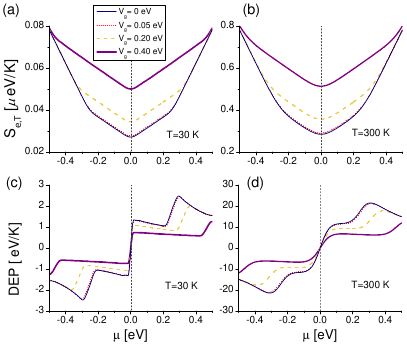}
\caption{Electronic entropies (a)-(b) and differential entropy per particle (c)-(d) as a function of the chemical potential for AAA-stacked trilayer graphene. Left panels stand for $T=30$ K, right panels $T=300$ K. Vertical dashed lines indicate $\mu=0$.}\label{fig7}
\end{figure}

The EC effect (or entropy changes) for AAA-stacked TLG is presented in Fig.\ \ref{DeltaSAAA-ABC}(a). The $-\Delta S_{e,T}$ response is linear as a function of $T$, following the linear behavior of the electronic spectra near $E_{\text{F}}$ of Fig.\ \ref{fig2}. The EC effect is inverse for all $V_g$ and the entire temperature range from 0 to 300 K, and reaches a high value $\simeq -0.23$ $\mu$eV/K at room temperature for large gate voltage ($V_g=0.4$ eV). This result can be explained directly from the plots of the electronic entropy as a function of the chemical potential presented in Fig.\ \ref{fig7}(a) at $T=30$ K and Fig.\ \ref{fig7}(b) $T=300$ K. The electronic entropy at $T=30$ K almost preserves the DOS shape of Fig.\ \ref{fig2}(b) as the $L$ function slowly smears the DOS at low temperature, while at $T=300$ K the entropies are largely smoothed with one magnitude order higher than $T=30$ K because of higher $T$. The electronic entropy at $\mu=0$ and $V_{g}=0$ for both temperatures $T=30$ K and $T=300$ K is the smallest entropy with respect to the other entropies with larger $V_g$. This means $S_{e}(V_{g}=0,T) < S_{e}(V_g\ne 0,T)$, hence we always obtain $-\Delta S_{e,T}<0$ from Eq.\ \ref{deltaS}, then the AAA-stacked TLG cools down for all $V_g$ we have considered here.

The DEP as a function of $\mu$ is shown in Fig.\ \ref{fig7}(c) and \ref{fig7}(d) for the same temperatures as for the electronic entropy calculations. The DEP for AAA TLG its an odd function of $\mu$, that is $s(-\mu)=-s(\mu)$, and at $\mu=0$ vanishes for all gate fields. At low temperature in Fig.\ \ref{fig7}(c), the DEP shows a \textit{stairslike} shape, mainly determined by the slope $\left ( \partial S_{e}/\partial \mu \right )_T$, with maxima (minima) localized near the discontinuities of the electron (hole) regime of the electronic entropy. The states of the maxima and minima of the DEP are around the shifted Dirac cones of the electronic dispersion of Fig.\ \ref{fig2}(a), which do not contribute to the EC effect as we mention above. At room temperature in Fig.\ \ref{fig7}(d), the DEP is smoothed and one magnitude order higher than $T=30$ K, as also seen for the electronic entropy, demonstrating a general linear behavior for the electronic and thermodynamics quantities in AAA-stacked TLG. 
\subsection{ABC-stacked TLG}
\begin{figure}[ht]
\centering
\includegraphics[width=\linewidth]{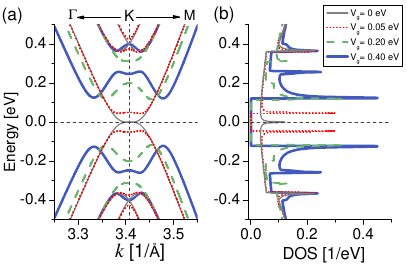}
\caption{(a) Band structure near the $K$ point, (b) DOS from the green triangle area in Fig.\ \ref{fig1}(d) for ABC-stacked trilayer graphene with diverse gate potential $V_g$ as indicated. Horizontal (vertical) dashed lines indicate $E_{\text{F}}=0$ eV ($K$ point). The hopping parameters are $\gamma_0=3.10$ eV, $\gamma_1=0.4$ eV and $\gamma_3=0.2$ eV \cite{bao2017stacking}.}
\label{fig3}
\end{figure}

Rhombohedral ABC-stacked trilayer graphene is one of the commonly stable crystal obtained in experimental procedures \cite{bao2011stacking,lui2011observation,bao2017stacking}, which can be considered as a zero-gap semiconductor material in the unperturbed case, and as a semiconductor when external electric-field potentials are applied \cite{aoki2007dependence}. This stacking geometry possesses inversion symmetry, but lacks mirror symmetry \cite{lui2011observation} as is schematized in Fig.\ \ref{fig1}(b). The TB Hamiltonian considering the hopping interactions in Fig.\ \ref{fig1}(b) is given by
\begin{equation}\label{HamiltonianABC}
\mathcal{\hat{H}}^{\mathrm{ABC}}(\pmb{k})= \begin{pmatrix} 
-V_g & \gamma_0 f_{0} & 0 & \gamma_3 f_{3} & 0 & 0 \\
\gamma_0 f_{0}^{*} & -V_g & \gamma{_1} f_{1}  & 0 & 0 & 0 \\
0 & \gamma{_1} f_{1}^{*} & 0 & \gamma_0 f_{0} & 0 & \gamma{_3} f_{3} \\
\gamma{_3} f_{3}^{*} & 0 & \gamma_0 f_{0}^{*} & 0 & \gamma{_1} f_{1} & 0 \\
0 & 0 & 0 & \gamma_1 f_{1}^* & V_g & \gamma_0 f_{0}\\
0 & 0 & \gamma{_3} f_{3}^{*} & 0 & \gamma_0 f_{0}^{*} & V_g
\end{pmatrix},
\end{equation}
where $f_3(\pmb{k})=f_0(k_x,k_y)e^{ick_z}$. The band structure and DOS from Eq.\ \ref{HamiltonianABC} considering $0\leq V_g \leq \gamma_1$ are plotted in Fig.\ \ref{fig3}(a) and Fig.\ \ref{fig3}(b) respectively. The electronic spectra preserve electron-hole symmetry around $E_{\text{F}}=0$ for all $V_g$. In the electronic structure there are valence and conduction bands dispersing according to the hoppings between the different atoms. For $V_g=0$ (thinnest gray lines), the lower conduction band and higher valence band touch at $E_{\text{F}}$ near the $K$ point, where the DOS shows a local maximum. 
Furthermore when $V_g=0$, two other pairs of bands shift away from $E_{\text{F}}$ and are degenerate, crossing near $E\simeq \pm 0.4$ eV because of the strongly interlayer coupling $\gamma_1$ between the direct bonded $B1$-$A2$ and $B2$-$A3$ dimers. The DOS increases at these crossing-band energies and presents a pair of local sharp peaks, not providing states to the EC effect for  $T > 900$ K. When $V_g$ is turned on, an energy gap between the lower electron (conduction band) and higher hole branch (valence band) opens at the $K$ point as well as in their vicinity because of the potential energy difference between the bottom and top graphene layers\cite{guinea2006electronic,latil2006charge,aoki2007dependence,koshino2009trigonal,avetisyan2010stacking,wu2011field}. The gaps non-linearly increase as $V_g$ increases \cite{wu2011field,wang2016first}, the DOS vanishes for the gap energies as expected, and shows two high symmetric peaks (van Hove singularities) near the valence and conduction band edges, whose states are mainly contributed by the carbon atoms of the top and bottom graphene layers \cite{wang2016first}. The bandgap edges for low-gate potential $V_g=0.05$ eV ($V_g=0.1$ eV) provide states to the EC effect for temperatures starting from $\sim 80$ K ($140$ K), while the states contribute for $T > 240$ K when high-gate fields $V_g=0.2, 0.4$ eV are applied. 

Figure \ref{DeltaSAAA-ABC}(b) shows the EC effect for ABC-TLG structure. Here we observe a non-linear direct response (heating) for the whole range of temperatures and gate potentials. This occurs as the reference electronic entropy [$S_e(V_g=0)$] is higher than the other gate-dependent entropies, because the states of the DOS evaluated at $D(E_{\text{F}}=0,V_g=0)$, are fully capturated by the $L$ function at low and high temperatures. We also have checked that there is not a direct-inverse transition [$-\Delta S_{e}(V_g,T)=0$] up to $T=500$ K. Moreover, an interesting effect occurs for temperatures below 80 K for all $V_{g}$, where the entropy changes are constant and do not present differences in the caloric response because of the presence of the band gap. As the temperature increases in Fig.\ \ref{DeltaSAAA-ABC}(b), the maxima of the entropy changes correspond to $-\Delta S_{e}(V_{g}=0.05\ \text{eV}, T=137\ \text{K})\simeq 0.14$ $\mu$eV/K and $-\Delta S_{e}(V_{g}=0.1\ \text{eV}, T=236\ \text{K})\simeq 0.20$ $\mu$eV/K. For high $V_g$ and $T>0$, $-\Delta S_{e}$ equally increases for $V_{g}=\gamma_{1}/2=0.2$ eV and $V_{g}=\gamma_{1}=0.4$ eV, reaching a caloric response of $\simeq 0.28$ $\mu$eV/K at room temperature. 

\begin{figure}[!h]
\centering
\includegraphics[width=\linewidth]{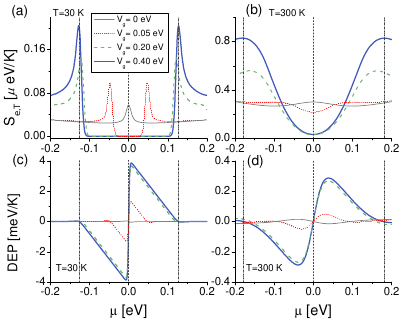}
\caption{Electronic entropies (a)-(b) and differential entropy per particle (c)-(d) as a function of $\mu$ for ABC-stacked trilayer graphene. Left panels stand for $T=30$ K, right panels $T=300$ K. Vertical dashed lines indicate $\mu=0$, and vertical dash-dotted lines in (a)-(b) at $\mu=\pm 0.127$ ($\mu=\pm 0.18$) eV highlight the maxima at $T=30$ ($T=300$) K for $S_{e,T}(V_g=0.4)$, which correlate with sign changes of the DEP in (c)-(d).}\label{fig8}
\end{figure}

These behaviors are consistent with the electronic entropy as a function of $\mu$ in Fig.\ \ref{fig8}(a) at $T=30$ K and Fig.\ \ref{fig8}(b) $T=300$ K. For both temperatures at $\mu=0$, the electronic entropy fulfills the condition $S_{e}(V_{g}=0,T) > S_{e}(V_{g}\neq 0,T)$, giving the direct EC response, then the ABC-TLG sample heats, opposite to the AAA-TLG case. At low temperature in the vecinity of $\mu=0$ in Fig.\ \ref{fig8}(a), the only entropy contributing to the EC effect is that for $V_{g}=0$, while the entropies when $V_{g} \neq 0$ vanish and do not provide states to the EC effect as expected. For temperatures lower than $80$ K, we have verified that $-\Delta S_{e}(V_g\neq 0)\simeq S_{e}(V_{g}=0)$ because of the band gap. At room temperature in Fig.\ \ref{fig8}(b), the electronic entropies are nearly five times larger than entropies at $T=30$ K, showing non null values for all $V_g$ in the gap region. The entropy is almost flat for $V_g=0$, while the other entropies have a parabolic shape as $V_g$ increases, with almost the same values for high gate-potential fields $V_g=0.2,0.4$ eV near $\mu=0$, as expected from the EC response in Fig.\  \ref{DeltaSAAA-ABC}(b). 

The DEP of ABC-stacked TLG also is an odd function of $\mu$ at low and high temperatures, as shown in Fig.\ \ref{fig8}(c) and Fig.\ \ref{fig8}(d) respectively. At $T=30$ K the peak value of $|s(\mu \simeq 0)|$ is three magnitude orders lower than that of AAA TLG, and increases as the gate field increases. 
As the EC effect in Fig.\ \ref{DeltaSAAA-ABC}(b) at $T=30$ K has not dependence on $V_g$, the DEP presents sizable dip-peak structures in the band gap zone, with large slopes in the vicinity of $\mu=0$ for $V_g \neq 0$, while $s(\mu)=0$ for $V_g=0$ except near $E_{\text{F}}$. The chemical potential dependence for $s(\mu)$ gives zeros when $S_{e,T}(\mu)$ reach maxima, as indicated with vertical dash-dotted lines particularly for $V_g=0.4$ eV. At low temperatures, zeros of $s(\mu)$ and maxima of $S_{e,T}(\mu)$ are located near the band edge Van Hove singularities of the electronic spectra in Fig.\ \ref{fig3}. This behavior is similar to the thermopower of biased bilayer graphene \cite{hao2010thermopower} and other gapped 2D materials such as germanene under an external electric field \cite{grassano2018detection} and semiconducting dichalcogenides \cite{shubnyi2018entropy}. At room temperature, the DEP in Fig.\ \ref{fig8}(d) decreases one magnitude order along with smoother resonances, showing a sinelike shape about $E_{\text{F}}$ for all $V_g \neq 0$. Near $\mu=0$ the slopes decrease as the gate potential decreases. The slope reverses for $V_g=0$ because of the maximum of $S_{e,T}(\mu=0)$ instead a minimum. These results are in agreement with the EC response in Fig.\ \ref{DeltaSAAA-ABC}(b). 
\subsection{ABA-stacked TLG}
Trilayer graphene with Bernal-ABA stacking is the most stable geometry as resembles a graphitelike atomic structure \cite{aoki2007dependence}. The ABA TLG possesses mirror symmetry with respect to the middle graphene layer. This structural symmetry is broken in the presence of electrostatic potentials \cite{guinea2006electronic,koshino2009gate}, driving to semimetallic character with tunable overlap of the electronic bands \cite{aoki2007dependence,craciun2009trilayer,wang2016first}, as well as controllable gaps at $K$ point and near it \cite{avetisyan2009electric,wu2011field,wang2016first}. The TB Hamiltonian considering all hoppings in Fig.\ \ref{fig1}(c) is given by
\begin{equation}\label{HamiltonianABA}
\mathcal{\hat{H}}^{\mathrm{ABA}}(\pmb{k})= \begin{pmatrix} 
-V_g & \gamma_0 f_{0} & \gamma_4 f_{3} & \gamma_3 f_{3} & \gamma_2 f_{1} & 0 \\
\gamma_0 f_{0}^{*} & -V_g & \gamma{_1} f_{1}  & \gamma_4 f_{3} & 0 & \gamma_5 f_{1} \\
\gamma_4 f_{3}^{*} & \gamma{_1} f_{1}^{*} & 0 & \gamma_0 f_{0} & \gamma_4 f_{3}^{*} & \gamma{_1} f_{1} \\
\gamma{_3} f_{3}^{*} & \gamma_4 f_{3}^{*} & \gamma_0 f_{0}^{*} & 0 & \gamma{_3} f_{3}^{*} & \gamma_4 f_{3}^{*} \\
\gamma_2 f_{1}^{*} & 0 & \gamma_4 f_{3} & \gamma_3 f_{3} & V_g & \gamma_0 f_{0}\\
0 & \gamma{_5} f_{1}^{*} & \gamma_1 f_{1}^{*} & \gamma_4 f_{3} & \gamma_0 f_{0}^{*} & V_g
\end{pmatrix}.
\end{equation}
We analyze first a simple case for ABA TLG, labeled nearest-layer (NL) TB model, including hopping parameters $\gamma_0$, $\gamma_1$ and $\gamma_3$, with $\gamma_2=\gamma_4=\gamma_5=0$ from spectroscopy experiments \cite{bao2017stacking}. We take different values of the external gate potential in the limit $0\leq V_g\gtrsim \gamma_1$ for the electronic spectra and EC calculations. 

\begin{figure}[!h]
\centering
\includegraphics[width=\linewidth]{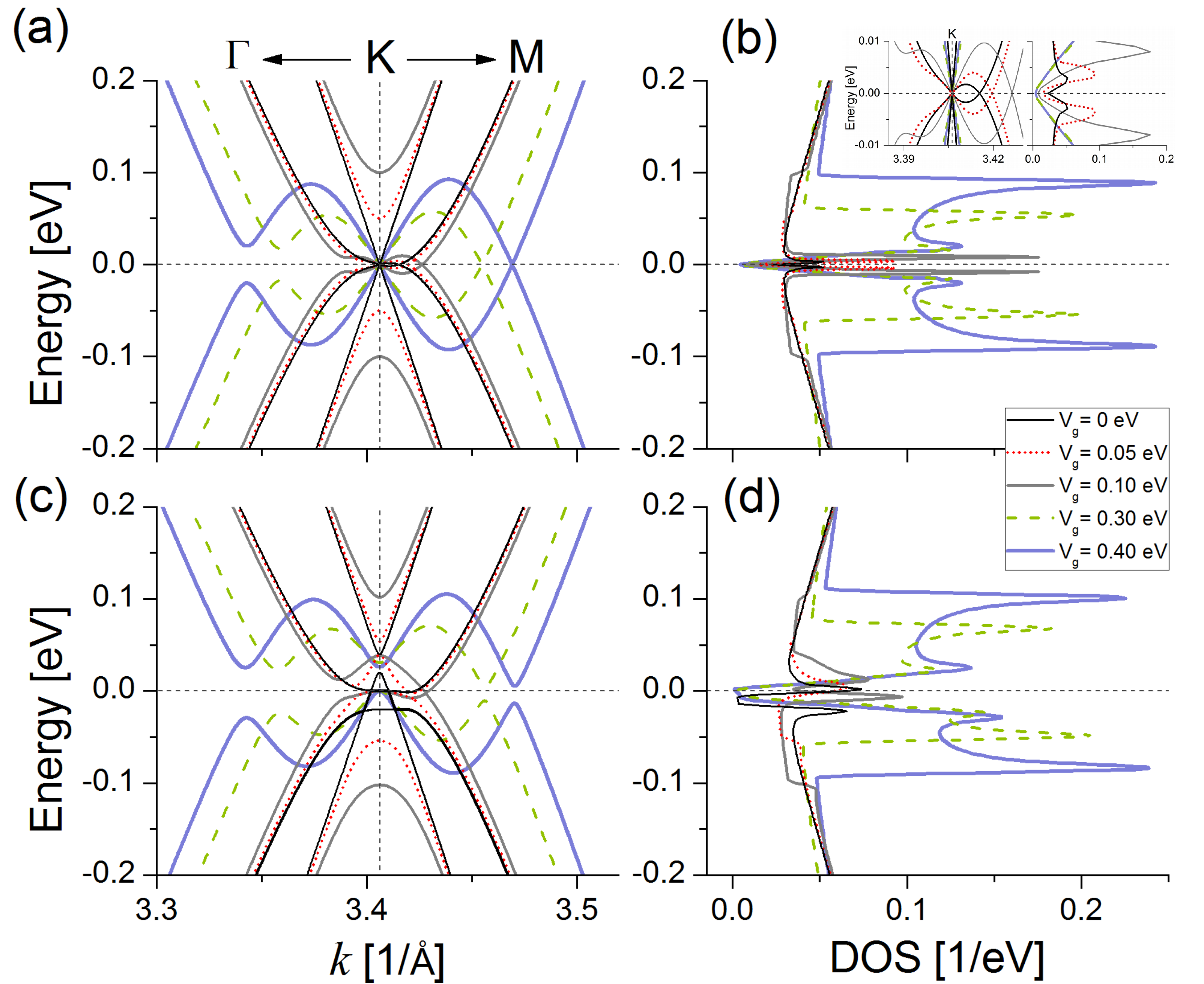}
\caption{Electronic spectra for ABA-stacked trilayer graphene near the Fermi level using nearest layer NL (a)-(b) and next-nearest layer NNL (c)-(d) TB models. Left panels show the electronic band structure about $K$ point in the Brilloin zone. Right panels ilustrate the density of states for the triangular area enclosed by green lines in Fig.\ \ref{fig1}(d). (a)-(b) $\gamma_0=3.15$ eV, $\gamma_1=0.39$ eV and $\gamma_3=0.25$ eV \cite{bao2017stacking}. (c)-(d) $\gamma_0=3.16$ eV, $\gamma_1=0.39$ eV, $\gamma_2=-0.020$ eV, $\gamma_3=0.315$ eV, $\gamma_4=0.044$ eV and $\gamma_5=-0.04$ eV \cite{koshino2009trigonal}. Horizontal (vertical) dashed line indicates $E_{\text{F}}$ ($K$ point). Inset in (b) shows a zoomed area for the spectra of the NL TB model around $E_{\text{F}}=0$.}
\label{fig4}
\end{figure}

Figure \ref{fig4}(a) show the electronic band structure around $K$ point, Fig.\ref{fig4}(b) correspond to the DOS. Both quantities are shown in the vicinity of $E_{\text{F}}=0$. The energy spectra are symmetric about zero energy with and without the presence of the gate field, preserving electron-hole symmetry \cite{koshino2009gate}. For $V_g=0$ (solid black lines), two bands mimic the $k$-linear electronic dispersion of monolayer graphene at the $K$ point, and one pair of parabolic bands overlap along $KM$ $k$-axis near $E_{\text{F}}$. The whole unperturbed bands ($V_g=0$) give a superposition of states of monolayerlike and Bernal-bilayerlike graphene states \cite{guinea2006electronic,partoens2006graphene,partoens2007normal}. The DOS for $V_g=0$ shows a sharp minimum at $E_{\text{F}}$ with two symmetric peaks surrounded it [see inset of Fig.\ref{fig4}(b)]. The major contribution of the edge peaks its from the AB-bilayerlike electron and hole band edges, whose states are fully captured by the $L$ function for $T>10$ K. 

The gate field breaks mirror reflection symmetry with respect to the central layer and mixes the linear and parabolic bands in the region of $E_{\text{F}}$ \cite{guinea2006electronic,koshino2009gate}, so that when $V_g=0.05$ eV (red dotted lines) and $V_g=0.1$ eV (gray solid lines), the monolayerlike linear bands are shifted from $E_{\text{F}}$, and the bilayerlike bands remain almost unperturbed. The DOS for these cases presents a minimum at $E_{\text{F}}$, and shows large electron and hole edge peaks as more states are overlapped near $E_{\text{F}}$, where the electronic states contribute to the EC effect for $T>30$ K. When the gate potential increases comparable to $\gamma_3 \sim V_g =0.3$ eV (dashed green lines) and $\gamma_1 \simeq V_g =0.4$ eV (solid purple lines), linear dispersion is seen at the $K$ point as well as along the $KM$ direction near $E_{\text{F}}$, while a gap opens along $K\Gamma$ as $\gamma_3$ is non zero \cite{avetisyan2009electric,bao2017stacking}. For these large-gate field potentials, the DOS almost vanishes at $E_{\text{F}}$ and shows one pair of symmetric peaks centered near the gap edges, contributing to the EC response when $T \geq 70$ K. Other pair of peaks are centered in the vicinity of the minima of the highest valence band and maxima of the lowest conduction band, not providing states to the EC effect for $T < 170$ K.

When all the interactions between the atoms are taken into account in the ABA TLG system, labeled as next-nearest layer (NNL) TB model, the energy spectra are no longer electron-hole symmetric about $E_{\text{F}}$ as in the NL case because of the additional electron hoppings. However, the full band structure for $V_g=0$ can still be considered as a combination of the band diagram of a single graphene layer and that of AB-bilayer graphene as shown in Fig.\ \ref{fig4}(c). At $K$ point, the valence and conduction parabolic bands open a gap near $E_{\text{F}}$ for $V_g=0$, where the DOS shows a flat minimum in the hole zone ($E_{\text{F}} \lesssim 0$), Fig.\ \ref{fig4}(d). The electron and hole peak edges from the AB-bilayerlike bands in this case, provide states to the EC effect for higher temperatures than in the NL case, $T>60$ K. 
When $V_g=0.05$ eV (dotted red lines), the gate potential causes anticrossing of the parabolic bands, similar as seen when negative and positive low-charged gates act on a suspended ABA TLG \cite{avetisyan2009electric}. At $E_{\text{F}}=0$ the DOS increases, showing nearly a maximum instead of a minimum as in the NL case. The electronic states belonging to this maximum are fully captured by the $L$ function for $T>30$ K. For $V_g=0.1$ eV, the bands are similar to the NL model, but opening a gap at the $K$ point, and showing a highly antisymmetric DOS, whose states near $E_{\text{F}}$ contribute to the EC effect for $T>60$ K. Large gate potentials $V_g=0.3$ eV and $V_g=0.4$ eV open several gaps at the $K$ point as well as along both $K\Gamma$ and $KM$ $k$-axis, similar to the effect of gate-induced high charge density on ABA TLG \cite{avetisyan2009electric}. The DOS for these high $V_g$ values show minima at zero energy, as in the band structure there are allowed states only near the $K$ point, contributing to the EC response for $T \geq 30$ K.              

\begin{figure}[!h]
\centering
\includegraphics[width=\linewidth]{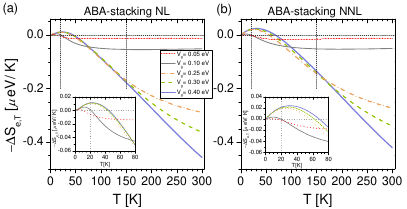}
\caption{The isothermal entropy changes $-\Delta S_{e,T}$ as a function of temperature $T$ for different gate potentials as indicated. (a) ABA-stacked trilayer graphene considering NL hoppings, (b) ABA-stacked trilayer graphene with NNL hoppings. The chemical potential is set to $\mu=0$ eV. Horizontal dashed line indicates $-\Delta S_{e,T}=0$, while vertical dashed lines stand for $T=20$ K and $T=150$ K, where the entropies are calculated as a function of $\mu$ in Fig.\ \ref{fig10}. Insets show a zoom for each ABA-EC response at relatively low temperatures.}\label{fig6}
\end{figure}

In Fig.\ \ref{fig6}(a) and \ref{fig6}(b), we present respectively the EC response for both the NL-ABA and NNL-ABA TB model. Both cases are qualitatively similar as they show a dual behavior seen as a combination of the direct and inverse EC effect, whose gate-dependent inversion $-\Delta S_{e,T}(V_g)=0$ occurs at relatively low temperatures. Moreover, noticeable differences are observed for the entropy changes depending on the magnitude of the applied gate field in each ABA-TB case. We observe a small direct response zone (heating) for low gate-fields $V_{g}=0.05$ (dotted red lines) and $V_{g}=0.1$ eV (solid gray lines) in both ABA-TB models, with maxima $-\Delta S_{e} \leq 0.005$ $\mu$eV/K at $T\leq 12$ --see insets of Fig.\ \ref{fig6}. As temperature increases for these low-gate field values, the entropy changes become inverse for $T \leq 21$ and remain constant as temperature increases, cooling down up to $T=300$ K.
 
The EC effect for high-gate fields $V_{g}=0.25,0.3,0.4$ eV is very different, as the direct caloric response zone (the maxima and inversion entropy changes) occurs at higher temperatures as compared to low-gate field values. For the ABA-NL case in Fig.\ \ref{fig6}(a), the maxima of the entropy changes are almost constant, reaching $-\Delta S_{e} \simeq 0.011$ $\mu$eV/K at $T \simeq 23$ K. The entropy inversion temperature also is nearly the same for all high $V_{g}$, occurring at $T\sim 45$. The entropy changes maxima for high-gate fields increase almost twice in the ABA-NNL model, with higher temperatures in the range $26\ \text{K} \leq T \leq 30$ K. This is because the $L$ function captures a maximum instead of a minimum of the reference DOS [$D(V_{g}=0)$] in the NNL-ABA model. Remarkably, the inversion temperature in the ABA-NNL model varies with high-gate fields, $56\ \text{K} \leq T \leq 69$ K. As the temperature increases, the entropy changes increase up to room temperature, cooling more for high gate field values. The inversion temperatures for the ABA-NNL system also can be obtained in zone II of Fig.\ \ref{fig11}(a), where the reference entropy (black line) crosses with all other entropies.  

\begin{figure}[!h]
\centering
\includegraphics[width=\linewidth]{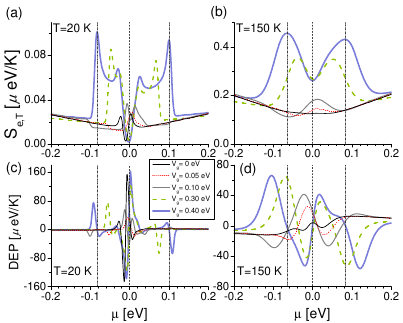}
\caption{Electronic entropies (a)-(b) and differential entropy per particle DEP (c)-(d) as a function of $\mu$ for ABA-stacked trilayer graphene considering the NNL-TB model. Left panels stand for $T=20$ K, right panels $T=150$ K. Vertical dashed lines indicate $\mu=0$ for all plots. Vertical dash-dotted lines at $\mu=-0.083$ eV and $\mu=0.1$ eV in (a); and at $\mu=-0.065$ eV, $\mu=0.083$ eV in (b) highlight the maxima for $S_{e,T}(V_g=0.4)$, which correlate with sign changes of the DEP in (c) and (d) respectively.}\label{fig10}
\end{figure}

\begin{figure}[!h]
\centering
\includegraphics[width=\linewidth]{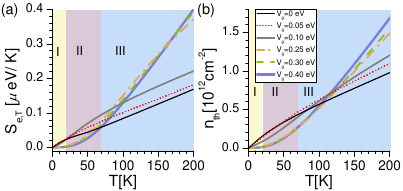}
\caption{(a) Electronic entropy $S_e(T)$, (b) density of thermally excited electrons $n_{th}(T)$ versus temperature calculated at $E_\text{F}=0$ in ABA-stacked TLG within the NNL-TB model for different gate potentials $V_g$. In (a), I: $S_e(V_g=0) \geq S_e(V_g\neq 0)$ , II: $S_e(V_g=0.05,0.1\ \text{eV})\geq S_e(V_g=0)\geq S_e(V_g=0.25,0.3,0.4\ \text{eV})$, III: $S_e(V_g=0)\leq S_e(V_g\neq 0)$. Zone I, II and III in (b) corresponds to the same temperature range as in (a).}\label{fig11}
\end{figure}

The direct-inverse response in ABA-TLG is captured by the electronic entropy as a function of $\mu$ in Fig.\ \ref{fig10}(a) and \ref{fig10}(b). We present here only the ABA-NNL plots as both ABA system EC results are equivalent. Figure \ref{fig10}(a) shows $S_{e,T}$ at $T=20$ K resembles the DOS shape of Fig.\ \ref{fig4}(d) with broken electron-hole symmetry. At $\mu=0$, the reference entropy for $V_{g}=0$ is lower than the entropy for $V_{g}=0.05$ eV, and larger than the entropies for all other gate field values, see Fig.\ \ref{fig11}(a). Consequently, the sample cools for $V_{g}=0.05$ eV, and heats for higher gate fields at $T=20$ K, as expected from the NNL-EC response in Fig.\ \ref{fig6}(b). When the temperature increases up to  $T=150$ K in Fig.\ \ref{fig10}(b), the electronic entropies are nearly four times the entropies at $T=20$ K, and are smoothed because of higher temperature. The reference entropy shows a flat line shape near $\mu=0$ and is smaller than all other entropies, indicating the system cools for all gate fields at $T=150$, as also seen from the EC response.

Figure \ref{fig10}(c) and \ref{fig10}(d) show the DEP for the ABA-NNL model at $T=20$ K and $T=150$ K respectively. $s(\mu)$ is not a symmetric function of $\mu$ because of broken electron-hole symmetry for all gate-potential fields. Near zero chemical potential and $V_g=0,0.3,0.4$ eV at 20 K, where the sample heats [see inset of Fig.\ \ref{fig6}(b)], the DEP shows large spikes, which can be attributed to the semimetal character and the imbalance between electron and hole carriers of the electron spectra in Fig.\ \ref{fig4}(c) and \ref{fig4}(d). There are two other smaller resonances when $\mu$ lies in the vicinity of the discontinuities of the electronic entropy, shifted to larger $\mu$ as $V_g$ increases. These entropy discontinuities at 20 K are located near the minima of the valence band and maxima of the conduction band of the electron dispersion. $s(\mu)$ decreases for low gate fields $V_g=0.05,0.1$ eV in the vicinity of $E_{\text{F}}$, near where the EC response displays a sign change at $\sim$21 K [see inset of Fig.\ \ref{fig6}(b)]. As the ABA TLG gets cool for all $V_g$ at 150 K [Fig.\ \ref{fig6}(b)], the DEP in Fig.\ \ref{fig10}(d) reduces by almost half near $\mu=0$, and the resonances are now spread out, mixed with the shifted ones close to $S(\mu)$ discontinuities.

Because the ABA-stacked TLG is capable to heats and cools within the same sample as a direct and inverse EC response are obtained in Fig.\ \ref{fig6}, it is interesting to know how the electronic entropy $S_{e}$ is related to the charge density at the microscopic level. We calculate the number of carriers thermally excited of ABA-stacked TLG through the NNL-TB model. We determine the density of excited electrons $n_{th}$ from the valence band to the conduction band of Fig.\ \ref{fig4}(c) as a function of temperature for selected values of the gate potential. In a similar way to Eq.\ \ref{numelectrons}, $n_{th}$ above the Fermi level at a given temperature is given by $n_{th}(V_g,T)=A^{-1} \int_{\mu}^{E_h} D(E,V_g) n_{\text{F}}(E,T,\mu) dE$, where $A$ is the unit cell area of the TLG, and we take $\mu=0$ as the Fermi energy at $T=0$. 

The temperature dependence of $S_e(T)$ and $n_{th}(T)$ is shown in Fig.\ \ref{fig11}(a) and Fig.\ \ref{fig11}(b) respectively. The fraction of excited electrons follows a very similar pattern as the electronic entropy because charge carriers contribute to the entropy by filling electronic states as temperature increases. In Fig.\ \ref{fig11}(b), electrons are excited at low temperature for low gate potentials $V_g=0.05,0.1$ eV, where the electronic dispersion and DOS of the parabolic bands present a negligible bandgap at $E_\text{F}=0$, see Fig.\ \ref{fig4}(c) and \ref{fig4}(d). As the gate potential increases, the energy gaps of ABA-TLG increase, and a higher temperature is needed to excite electrons from the valence band to the conduction band. This produces a suppression (non-linear behavior) of $n_{th}$ at low temperatures for high electric potentials. From $\sim 70$ K, $n_{th}$ linearly increases with temperature at high temperatures, which may correspond to a monotonic DOS as a function of energy, that is gapless parabolic electronic bands near $E_{\text{F}}$. Similar behavior for the density of excited electrons has been experimentally observed in graphene multilayers including AB-bilayer and ABA-TLG, correlated with a theoretical temperature-dependent gap approximation, where both procedures give account of a finite-temperature electronic phase transition \cite{nam2018family}. In our case, large gate-field values $V_{g}=0.25,0.3,0.4$ eV induced overlap and gaps in the electronic structure of ABA-TLG, which compete with temperature in the proximity of $E_{\text{F}}$. This causes the direct EC effect $-\Delta S_e > 0$ with maxima and inversion values located in zone II of Fig.\ \ref{fig11}(a), which correspond to the charge suppression zone II of Fig.\ \ref{fig11}(b), that is gaped electronic bands. On the contrary, the inverse EC effect $-\Delta S_e < 0$ is located in zone III of $S_e(T)$, where electrons are thermally excited in a linear way as a function of temperature, indicating gapless bands near $E_{\text{F}}$. 

\section{Conclusions}
Gated trilayer graphene structures show an electrocaloric response that depends on the stacking pattern and the electronic character they possess near the Fermi level. Trilayer graphene with AAA stacking remains metallic under the applied gate field, presenting large entropy changes as the gate potential increases, linearly cooling up to room temperature. In striking contrast, ABC-stacked trilayer graphene converts into a semiconductor when the gate field is turned on, non-linearly heating as temperature increases. The energy spectra of ABA-stacked TLG geometry present semimetallic as well as gapped character with gate voltage, cooling and heating in the caloric response in a linear and non-linear mixed way, giving a combination of AAA and ABC electrocaloric effect. We verified that gate-dependent thermally excited charge at $E_{\text{F}}=0$ for ABA-TLG can be directly linked to the electronic entropy as well as to the EC response, where charge suppression is related to the direct EC effect, and linearly excited electrons correspond to the inverse EC effect. 

The differential entropy per particle shows particular signatures for each TLG structure at low and high temperatures. Measuring this thermodynamic quantity could be used as a cheaper and novel experimental technique to identify the stacking pattern in trilayers graphene and-or similar layered 2D materials. It is worth mentioning that a new caloric effect based on the entropy per particle and EC effect connection can be obtained, see Appendix. We refer to this mechanism as \textit{charge-carrier caloric effect}, as the variable quantity are charge carriers instead of an external field. We propose this effect may be measured by varying the charge density and keeping constant the electric field in the sample, similar to a 2D electron gas experimental setup \cite{kuntsevich2015strongly}.     

\section{Acknowledgments}
N.C. acknowledges support from ANID Fondecyt Postdoctoral Grant No. 3200658, P.V. and O.N. acknowledge support from ANID PIA/Basal AFB18000. F.J.P. acknowledges support from ANID Fondecyt, Iniciaci\'on en Investigaci\'on 2020 grant No. 11200032, and the financial support of USM-DGIIE.

\appendix

\section*{Appendix: Charge-Carrier Caloric Effect}\label{appendix}

The connection between the electrocaloric effect and the differential entropy per particle $s$ defined in Sec.\ \ref{sec:DEP}, can be obtained by analyzing the Helmholtz free energy $F$ and internal energy $U$ of a general system that includes a variable number of particles $N$ in its formulation. 

First, we define the differential internal energy of the system as
\begin{equation}
\label{diffU}
dU = TdS + \textbf{E}\cdot d\textbf{P} + \mu dN,
\end{equation}
where the first term corresponds to heat $\delta Q = T dS$, with $S$ the total entropy; the second term is the definition of electric work $W=\int \textbf{E}\cdot d\textbf{P}$, where $\textbf{P}$ is the electric polarization and $\textbf{E}$ is the electric field. The last term $\mu dN$, where $\mu$ is the chemical potential, is the energy needed to add or remove one particle of the system. On the other hand, $F$ is defined as 
\begin{equation}
\label{Freenergy}
F = U-TS-\textbf{E}\cdot \textbf{P}, 
\end{equation}
and its differential is
\begin{equation}
\label{diffF}
dF = -S dT - \textbf{P} \cdot d\textbf{E} +\mu dN,
\end{equation}
where we have used Eq.\ \ref{diffU}. From Eq.\ \ref{diffF}, we can obtain the following Maxwell relations:
\begin{equation}
\label{maxwell1}
\left(\frac{\partial S}{\partial \textbf{E}}\right)_{T,N} = \left(\frac{\partial \textbf{P}}{\partial T}\right)_{\textbf{E},N},
\end{equation}
and 
\begin{equation}
\label{maxwell2}
\left( \frac{\partial S}{\partial N} \right)_{T,\textbf{E}} = -\left(\frac{\partial \mu}{\partial T}\right)_{\textbf{E},N}, 
\end{equation}
where the last equation is the same as presented in Eq.\ \ref{DEP}, but including the thermodynamic variable associated with the electric field, and the subscripts refer to variables held constant during partial differentiation.

To take into account the electronic entropy variations in our system, it is convenient to specify the total differential entropy, whose expression is given by
\begin{equation}
\label{totaldiffS}
dS = \left(\frac{\partial S}{\partial T}\right)_{\textbf{E},N} dT + \left(\frac{\partial S}{\partial \textbf{E}}\right)_{T,N} d\textbf{E} + \left(\frac{\partial S}{\partial N}\right)_{T,\textbf{E}} dN.
\end{equation}
If we calculate the entropy change in an isothermal process ($dT=0$) along a path of the form $S(\textbf{E}_{1}, N_{1})\rightarrow S(\textbf{E}_{2}, N_{1})\rightarrow S(\textbf{E}_{2}, N_{2})$, we obtain
\begin{equation}
\label{deltasisogeneral}
\Delta S_{T}=\int_{\textbf{E}_{1}}^{\textbf{E}_{2}} \left(\frac{\partial S}{\partial \textbf{E}}\right)_{T,N = N_{1}} d\textbf{E} + \int_{N_{1}}^{N_{2}} \left(\frac{\partial S}{\partial N}\right)_{T, \textbf{E} = \textbf{E}_{2}} dN.
\end{equation}
Employing the Maxwell relations of Eq.\ \ref{maxwell1} and Eq.\ \ref{maxwell2}, we finally obtain
\begin{equation}
\label{generalisothermicS}
\Delta S_{T} = \int_{\textbf{E}_{1}}^{\textbf{E}_{2}} \left(\frac{\partial \textbf{P}}{\partial T}\right)_{\textbf{E},N}  d\textbf{E} - \int_{N_{1}}^{N_{2}} \left(\frac{\partial \mu}{\partial T}\right)_{N, \textbf{E} = \textbf{E}_{2}} dN.
\end{equation}
This last equation occurs for an isothermal process in which the electric field and the number of particles vary in the process. The terms associated with the DEP corresponds to the integrand in the second term of Eq.\ \ref{deltasisogeneral} and Eq.\ \ref{generalisothermicS}.  
This means that the DEP is directly connected to the electrocaloric phenomena. The \textit{charge-carrier caloric effect} will correspond to the measurement of a process in which the electric field remains fixed, but a variation in the number of particles is present in the initial and final stages of the process. Hence, the pure caloric effect obtained from the DEP is given by
\begin{equation}
\Delta S_{T}^{\text{DEP}}=-\int_{N_{1}}^{N_{2}}\left(\frac{\partial \mu}{\partial T}\right)_{N, \textbf{E}} dN.
\end{equation} 
It is essential to highlight that when the number of particles remains fixed, the usual expression of the electrocaloric effect is recovered from Eq.\ \ref{generalisothermicS}
\begin{equation}
\Delta S_{T} = \int_{\textbf{E}_{1}}^{\textbf{E}_{2}} \left(\frac{\partial \textbf{P}}{\partial T}\right)_{\textbf{E},N}  d\textbf{E}.
\end{equation}
As the differential entropy per particle is an experimentally measurable quantity, we believe that the associated \textit{charge-carrier caloric effect} can be performed through indirect measurements, not requiring precision calorimetry. A similar procedure has been recently reported by a direct measurement of $\partial \mu / \partial T$, obtaining very small electric currents, caused by a periodic temperature modulation in a 2D diluted electronic system \cite{kuntsevich2015strongly}.

\bibliography{bib}
\email{natalia.cortesm@usm.cl}

\end{document}